\documentclass[conference]{IEEEtran}
\IEEEoverridecommandlockouts
\usepackage{cite}
\usepackage{multirow}
\usepackage{booktabs}
\usepackage{url}
\usepackage{amssymb} 
\usepackage{amsmath,amssymb,amsfonts}
\usepackage{algorithmic}
\usepackage{graphicx}
\usepackage{textcomp}
\usepackage{hyperref}  
\hypersetup{
  colorlinks=true,
  linkcolor=blue,     
  citecolor=green     
}
\usepackage{subcaption}
\usepackage{xcolor}
\def\BibTeX{{\rm B\kern-.05em{\sc i\kern-.025em b}\kern-.08em
    T\kern-.1667em\lower.7ex\hbox{E}\kern-.125emX}}
\begin{document}

\title{Enhanced SCanNet with CBAM and Dice Loss for Semantic Change Detection}

\author{\IEEEauthorblockN{R.M.A.M.B. Ratnayake}
\IEEEauthorblockA{\textit{Dept. of Electrical and Electronic Eng.} \\
\textit{University of Peradeniya}\\
Peradeniya, Sri Lanka \\
e19328@eng.pdn.ac.lk}
\and
\IEEEauthorblockN{W.M.B.S.K. Wijenayake}
\IEEEauthorblockA{\textit{Dept. of Electrical and Electronic Eng.} \\
\textit{University of Peradeniya}\\
Peradeniya, Sri Lanka \\
e19445@eng.pdn.ac.lk}
\and
\IEEEauthorblockN{D.M.U.P. Sumanasekara}
\IEEEauthorblockA{\textit{Dept. of Electrical and Electronic Eng.} \\
\textit{University of Peradeniya}\\
Peradeniya, Sri Lanka \\
e19391@eng.pdn.ac.lk}
\and
\IEEEauthorblockN{G.M.R.I. Godaliyadda}
\IEEEauthorblockA{\textit{Dept. of Electrical and Electronic Eng.} \\
\textit{University of Peradeniya}\\
Peradeniya, Sri Lanka \\
roshang@eng.pdn.ac.lk}
\and
\IEEEauthorblockN{H.M.V.R. Herath}
\IEEEauthorblockA{\textit{Dept. of Electrical and Electronic Eng.} \\
\textit{University of Peradeniya}\\
Peradeniya, Sri Lanka \\
vijitha@ee.pdn.ac.lk}
\and
\IEEEauthorblockN{M.P.B. Ekanayake}
\IEEEauthorblockA{\textit{Dept. of Electrical and Electronic Eng.} \\
\textit{University of Peradeniya}\\
Peradeniya, Sri Lanka \\
mpbe@eng.pdn.ac.lk }
}

\maketitle

\begin{abstract}
Semantic Change Detection (SCD) in remote sensing imagery requires accurately identifying land-cover changes across multi-temporal image pairs. Despite substantial advancements, including the introduction of transformer-based architectures, current SCD models continue to struggle with challenges such as noisy inputs, subtle class boundaries, and significant class imbalance. In this study, we propose enhancing the Semantic Change Network (SCanNet) by integrating the Convolutional Block Attention Module (CBAM) and employing Dice loss during training. CBAM sequentially applies channel attention to highlight feature maps with the most meaningful content, followed by spatial attention to pinpoint critical regions within these maps. This sequential approach ensures precise suppression of irrelevant features and spatial noise, resulting in more accurate and robust detection performance compared to attention mechanisms that apply both processes simultaneously or independently. Dice loss, designed explicitly for handling class imbalance, further boosts sensitivity to minority change classes. Quantitative experiments conducted on the SECOND dataset demonstrate consistent improvements. Qualitative analysis confirms these improvements, showing clearer segmentation boundaries and more accurate recovery of small-change regions. These findings highlight the effectiveness of attention mechanisms and Dice loss in improving feature representation and addressing class imbalance in semantic change detection tasks.

\end{abstract}

\begin{IEEEkeywords}
Semantic change detection, Remote sensing imagery, CBAM, Dice Loss

\end{IEEEkeywords}

\section{Introduction}
\begin{figure*}[htbp]
    \centering
    \includegraphics[width=0.8\textwidth]{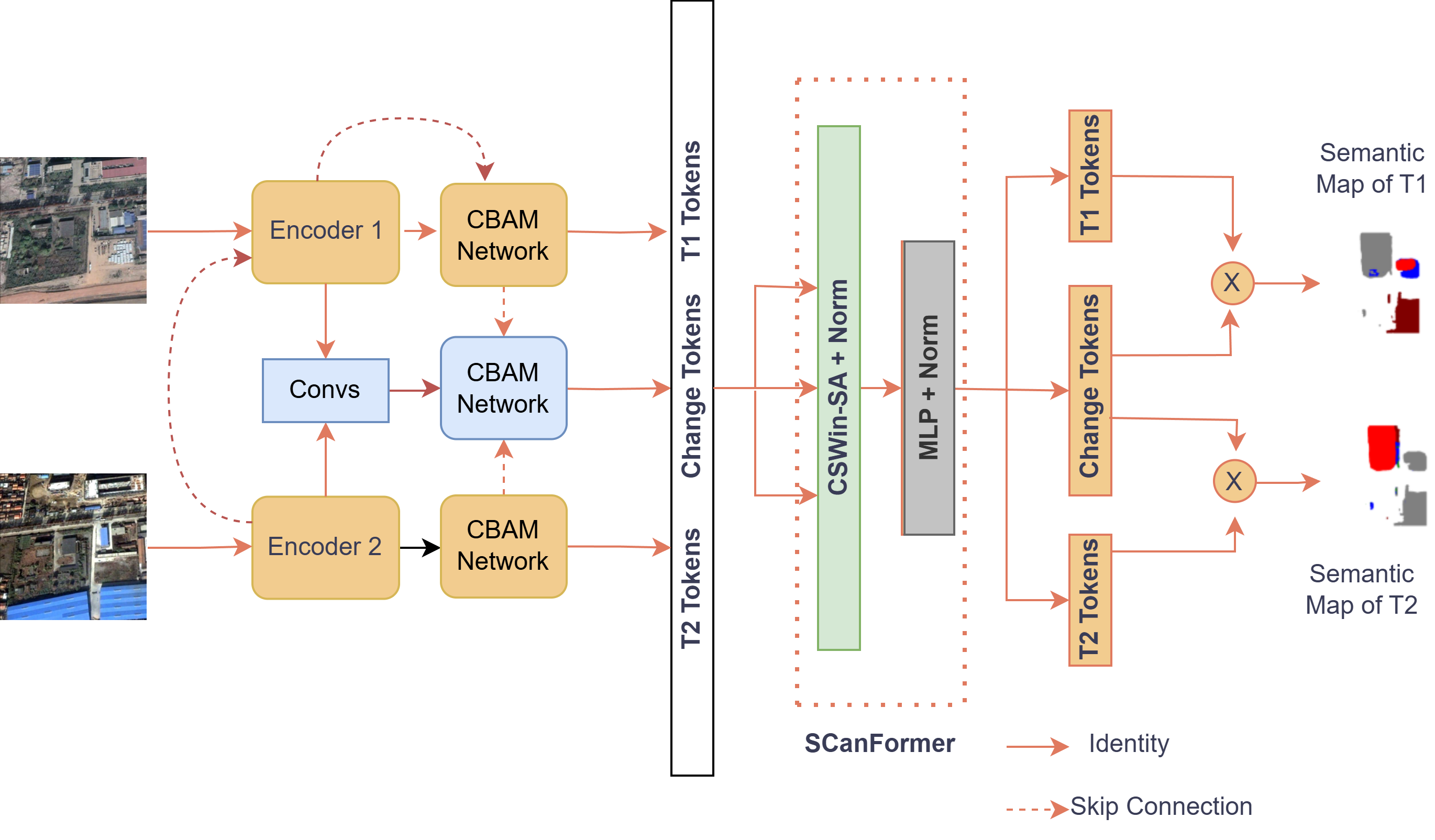}
    \caption{Proposed Enhanced SCanNet Architecture}
    \label{fig:overall}
\end{figure*}

Change Detection (CD) is an increasingly prominent area of research in the field of Remote Sensing which can be understood as the identification and characterization of change in Multi Temporal Images\cite{A_Zhu_2024}. In other words, CD involves analyzing two or more images of the same geographical area taken at different time points to determine the areas which have changed, and in certain cases, the nature of the change.

Due to the importance of analyzing land cover dynamics, CD finds many practical applications in a wide variety of fields. By leveraging satellite imagery, CD enables the analysis of temporal changes in geographical areas at significantly lower costs compared to traditional ground-based methods. This makes CD valuable in various fields, including urban development, environmental monitoring, agricultural assessment, and military surveillance\cite{C_Selvaraj_2022}.

Based on the level of detail in the output, CD Algorithms can be classified into several categories. Binary Change Detection (BCD) only determines whether a change has occurred between time points. In contrast, Multiple Change Detection (MCD) categorizes changes into multiple classes (e.g., urban expansion, deforestation, agricultural shifts). Semantic Change Detection (SCD) provides the highest level of detail, generating Land Cover/ Land Use (LCLU) maps at each time point in addition to a map detailing the changed areas. These LCLU maps label the land cover classes before and after the change, enabling a more comprehensive understanding of the nature and implications of changes, an essential feature for many practical applications \cite{A_Zhu_2024}.

However, due to the complexity and inherent variability of remote sensing imagery, Change Detection (CD) algorithms face several technical challenges. One major challenge is image variability: even identical features may appear significantly different across images due to variations in lighting conditions, sensor noise, atmospheric effects, and seasonal differences \cite{A_Zhu_2024}. This makes it difficult for algorithms to distinguish true changes from irrelevant variations. Another critical issue is image registration \cite{A_Zhou_2024}, as multi-temporal images are often slightly misaligned when captured at different times, leading to false detections. To address these challenges, incorporating both global and local contextual information has become essential in the design of robust CD algorithms.

A significant challenge affecting data-driven approaches is class imbalance in widely available CD datasets \cite{A_Zhu_2024}. In such cases, the distribution of class labels is heavily skewed, with a much smaller proportion of changed areas compared to unchanged ones. This imbalance makes it difficult for learning algorithms to accurately detect changes, as they tend to be biased toward the majority class, often leading to poor generalization and reduced performance on minority regions \cite{C_Leichtle_2017}.

In the early days of CD several classical algorithms were proposed that focus on pixel-based algorithms as well as statistical and signal processing approaches. Some common methods include Image Differencing, Image Ratioing \cite{D_Bai_2023}, Change Vector Analysis (CVA)\cite{U_Saha_2019}. Several Machine Learning methods such as Support Vector Machine (SVM) \cite{A_Bovolo_2008}, Decision Trees\cite{A_IM_2005}, Markov Random Fields and Conditional Random Fields\cite{G_Vemulapalli_2016,D_Bai_2023} have also been proposed. However, all these methods struggle with encapsulating the context and are susceptible to many of the challenges of CD. 

On the other hand, the use of deep learning has dominated the current CD landscape\cite{C_Chen_2024,D_Bai_2023} as it is better equipped to deal with the aforementioned technical challenges. In particular, Convolution Neural Network (CNN) based architectures have had significant success\cite{R_Li_2021,A_Jonasova_2023} due to their ability to capture local context effectively and their ability to learn complex and non-linear relationships. These models have significantly improved the accuracy and robustness of CD predictions\cite{C_Chen_2024}.

More recently, the integration of attention mechanisms \cite{J_Ding_2024,S_Liu_2021} into CD architectures has led to further performance gains. Attention modules enhance the model’s ability to capture global context by modeling long-range dependencies and complex spatial relationships, an essential capability in CD tasks, where both local and global information contribute meaningfully to accurate change interpretation\cite{C_Chen_2024}.

Furthermore, the most common CD architectures adopt an encoder–decoder backbone or one of its variants. The encoder transforms the incoming images into a latent space whereas the decoder transforms the latent space into the desired outputs, which are the Change Maps or LCLU Maps \cite{C_Chen_2024}. Often there is a fusion block that merges the latent feature embeddings of both images to generate the final change map. Frequently implemented in a Siamese configuration \cite{A_Yuan_2022}, these models feed each temporal image through identical, weight‑sharing encoder streams. To capture both local detail and global context, convolutional layers\cite{R_Li_2021} , attention mechanisms \cite{J_Ding_2024}, and Visual State Space (VSS)\cite{C_Chen_2024} blocks are interleaved throughout the network.

On the contrary, the possibility of using Convolutional Block Attention Module (CBAM) in CD has not been investigated in the existing literature. CBAM \cite{C_Woo_2018} allows the network to fuse together information from different stages of latent representation giving the network the ability to focus more on important channels as well as spatial features. 

In addition, Dice Loss \cite{G_H._2017} is a loss function designed to reduce the effect of class imbalance on training data, which is a key challenge in CD. Despite its relevance, thorough investigation as to its effectiveness as loss functions specifically for training CD algorithms, has received limited attention in the literature.

Identifying these research gaps, the contributions of the present work are as follows.

\begin{enumerate}
    \item Demonstrate the improvement of performance of CD architectures with the introduction of CBAM into the architecture.
    \item Demonstrate that Dice Loss improves Performance of the CD Algorithm
\end{enumerate}

\begin{figure*}[t]
  \centering
  \includegraphics[width=0.7\linewidth]{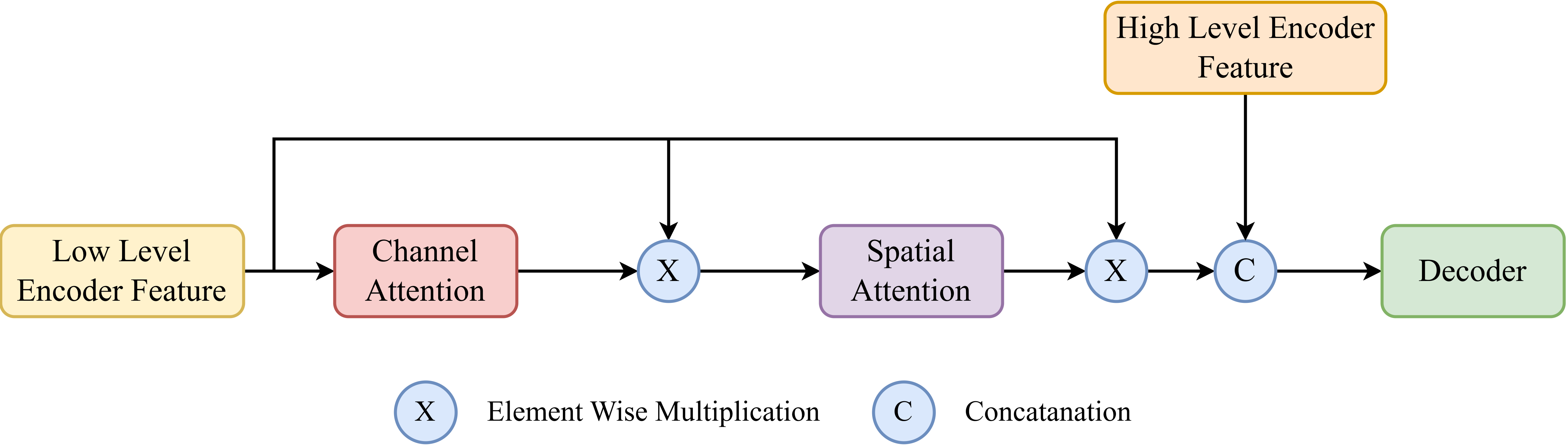}
  \caption{Proposed CBAM enhanced Decoder Block}
  \label{fig:cbam}
\end{figure*}
\section{Methodology}
\subsection{Semantic Change Network (SCanNet) Overview}

Our work builds upon the SCanNet framework of \cite{J_Ding_2024}, a hybrid CNN–Transformer tailored for SCD. SCanNet uses a Triple Encoder–Decoder (TED) backbone: two weight‑sharing CNN encoders extract pre‑ and post‑image semantic features independently, while a third encoder isolates change‐specific cues. Each encoder’s output is decoded through symmetric branches, ensuring high‐resolution segmentation maps that disentangle unchanged semantics from genuine changes.

However, purely convolutional TI pipelines struggle to model long‐range temporal interactions. To bridge this “semantic gap,” SCanNet’s SCanFormer module applies CSWin attention in a flattened token space, enabling each pixel to attend dynamically to both its own temporal counterpart and neighboring regions across time. This joint “from‑to” attention captures subtle transitions—e.g., gradual urban growth or seasonal vegetation shifts—that standard local convolutions may miss.

To further regularize training, SCanNet imposes two semantic constraints, (1) Pseudo‑label loss on no‑change regions, which uses high‑confidence network predictions to reinforce stable semantics and prevent drift, (2) Consistency loss on unchanged pixels, encouraging their semantic embeddings to remain similar across time while penalizing any residual similarity in truly changed areas.  

Together, these design choices—dual encoders, cross‑temporal attention, and targeted semantic losses—form a cohesive architecture \cite{J_Ding_2024} that excels at separating genuine land‑cover transitions from common remote‑sensing nuisances such as lighting, registration errors, and seasonal effects.

\subsection{CBAM Integration within SCanNet}
To enhance feature representations, we integrate CBAM into the decoder branches of SCanNet. CBAM sequentially applies channel and spatial attention\cite{C_Woo_2018}, refining feature maps by highlighting essential channels and spatial regions crucial for accurate change detection. The encoder is based on a standard ResNet-34 backbone, pre-trained on ImageNet, which extracts multi-scale features at different depths. The decoder integrates these features using skip connections and an attention-guided fusion mechanism\cite{D_He_2016}. Figure~\ref{fig:cbam} illustrates the decoder block integrating CBAM.

Effective integration of low-level spatial details and high-level semantic features is achieved through our CBAM Module, which combines Channel Attention (CA), Spatial Attention (SA), and concatenation-based fusion.

\subsubsection{Channel Attention (CA)}
Given an input feature map $x \in \mathbb{R}^{c\times H\times W}$, CA captures inter-channel dependencies via global average and max pooling operations, processed through a shared multilayer perceptron (MLP),

\begin{align}
    M_c &= \sigma(\text{MLP}(\text{AvgPool}(x)) + \text{MLP}(\text{MaxPool}(x))) \\
    x' &= x \odot M_c
\end{align}

\subsubsection{Spatial Attention (SA)}
SA refines feature maps by emphasizing informative spatial regions, computed from mean and max pooled channel information convolved with a $7 \times 7$ kernel:

\begin{align}
    M_s &= \sigma(f^{7 \times 7}([\text{Mean}(x');\text{Max}(x')])) \\
    x'' &= x' \odot M_s
\end{align}

\subsubsection{Attention Fusion Model}
The Attention Fusion module merges low-level encoder features ($x_{low}$) and high-level decoder features ($x_{high}$):

\begin{align}
    x_{low}^{att} &= x_{low} \odot M_c \odot M_s \\
    \text{fused} &= \text{Concat}(x_{high}, x_{low}^{att})
\end{align}

This fusion selectively enhances essential spatial details and semantic features, improving segmentation accuracy.

\subsection{Encoder-Decoder Feature Alignment}
Spatial mismatches between encoder and decoder features are resolved using bilinear upsampling. Each stage of the decoder integrates high-level features ($x_{high}$) with the corresponding low-level features ($x_{low}$) from the encoder using attention-based fusion (as illustrated in Figure~\ref{fig:cbam}) and skip connections. This design helps retain fine-grained spatial details while enriching the representation with high-level semantic information. The overall architecture is demonstrated in Figure~\ref{fig:overall}.

\begin{table*}[t!]
  \centering
  \caption{Quantitative results of the ablation study for the proposed techniques.}
  \label{tab:ablation}
  \resizebox{2\columnwidth}{!}{
  \begin{tabular}{l|cc|cccc}
    \toprule
    \multirow{2}{*}{Methods} 
      & \multicolumn{2}{c|}{Proposed Techniques} 
      & \multirow{2}{*}{OA (\%)} 
      & \multirow{2}{*}{$F_{scd}$ (\%)} 
      & \multirow{2}{*}{mIoU (\%)} 
      & \multirow{2}{*}{SeK (\%)} \\
    \cmidrule(lr){2-3}
      & CBAM & $\mathcal{L}_{\text{Dice}}$ & & & & \\
    \midrule
    TED~\cite{J_Ding_2024} 
      &  &  & 87.39 & 61.59 & 72.49 & 22.17 \\
    SCanNet~\cite{J_Ding_2024} 
      &  &  & 87.86 & 63.66 & 73.42 & 23.94 \\
    SCanNet + CBAM (Proposed)
      & \checkmark &  & 87.98 & 63.81 & 73.51 & 24.11 \\
    SCanNet + CBAM + $\mathcal{L}_{\text{Dice}}$ (Proposed) 
      & \checkmark & \checkmark & \textbf{88.12} & \textbf{64.31} & \textbf{73.63} & \textbf{24.25} \\
    \bottomrule
  \end{tabular}
  }
\end{table*}

\subsection{Loss Functions}
To address class imbalance and enhance boundary delineation, we augment SCanNet's original cross-entropy and pseudo-label losses with Dice Loss\cite{G_H._2017}. Additionally, we retain the temporal semantic consistency loss to leverage intrinsic bi-temporal coherence\cite{J_Ding_2024}.

\subsubsection{Semantic Loss for Changed Areas}
We define semantic losses for changed pixels $\Omega_c$,

\begin{align}
L_{\text{CE}}^{\text{sem}} &= -\sum_{i=1}^{2} [L_i\log Y_i^s + (1 - L_i)\log(1 - Y_i^s)], \label{eq:Lsem_CE} \\
L_{\text{Dice}}^{\text{sem}} &= 1 - \frac{2\sum_{i\in\Omega_c} Y_i^s L_i}{\sum_{i\in\Omega_c} Y_i^s + \sum_{i\in\Omega_c} L_i}, \label{eq:Lsem_Dice}
\end{align}

where $Y_i^s$ are predicted probabilities.

\subsubsection{Pseudo-Label Loss for Unchanged Areas\cite{J_Ding_2024}}
For unchanged pixels $\Omega_u$, we define pseudo-label losses,

\begin{align}
L^{\text{psd}} &= -\sum_{i=1}^{2}[\tilde{L}\log Y_i^s + (1 - \tilde{L})\log(1 - Y_i^s)], \label{eq:Lpsd_CE}
\end{align}

\subsubsection{Semantic Consistency Loss\cite{J_Ding_2024}}
We enforce that unchanged areas produce consistent semantics across time while changed areas differ, using the ground‐truth change map \(L_c\),
\begin{equation}
L_{\text{sc}}
= \sum_{p\in\Omega}\begin{cases}
1 - \cos\bigl(Y^s_{1,p},Y^s_{2,p}\bigr), & L_{c,p}=1\\
\cos\bigl(Y^s_{1,p},Y^s_{2,p}\bigr), & L_{c,p}=0
\end{cases}
\label{eq:Lsc}
\end{equation}
where \(\cos()\) is the cosine similarity of the semantic probability vectors.

\subsubsection{Overall Objective}
Integrating these components, our overall training objective becomes,

\begin{equation}
\begin{aligned}
L_{\text{total}}
&= \alpha\bigl(L_{\text{CE}}^{\text{sem}} + \lambda_1 L_{\text{Dice}}^{\text{sem}}\bigl) + \beta L_{\text{CE}}^{\text{psd}} + \gamma\,L_{\text{sc}}
\end{aligned}
\end{equation}

with hyperparameters optimized via cross-validation to balance semantic accuracy, overlap precision, and temporal consistency effectively.

\section{DATASET DESCRIPTION AND EXPERIMENTAL SETTINGS}
\begin{figure*}[t]
    \centering
    \includegraphics[width=1\linewidth]{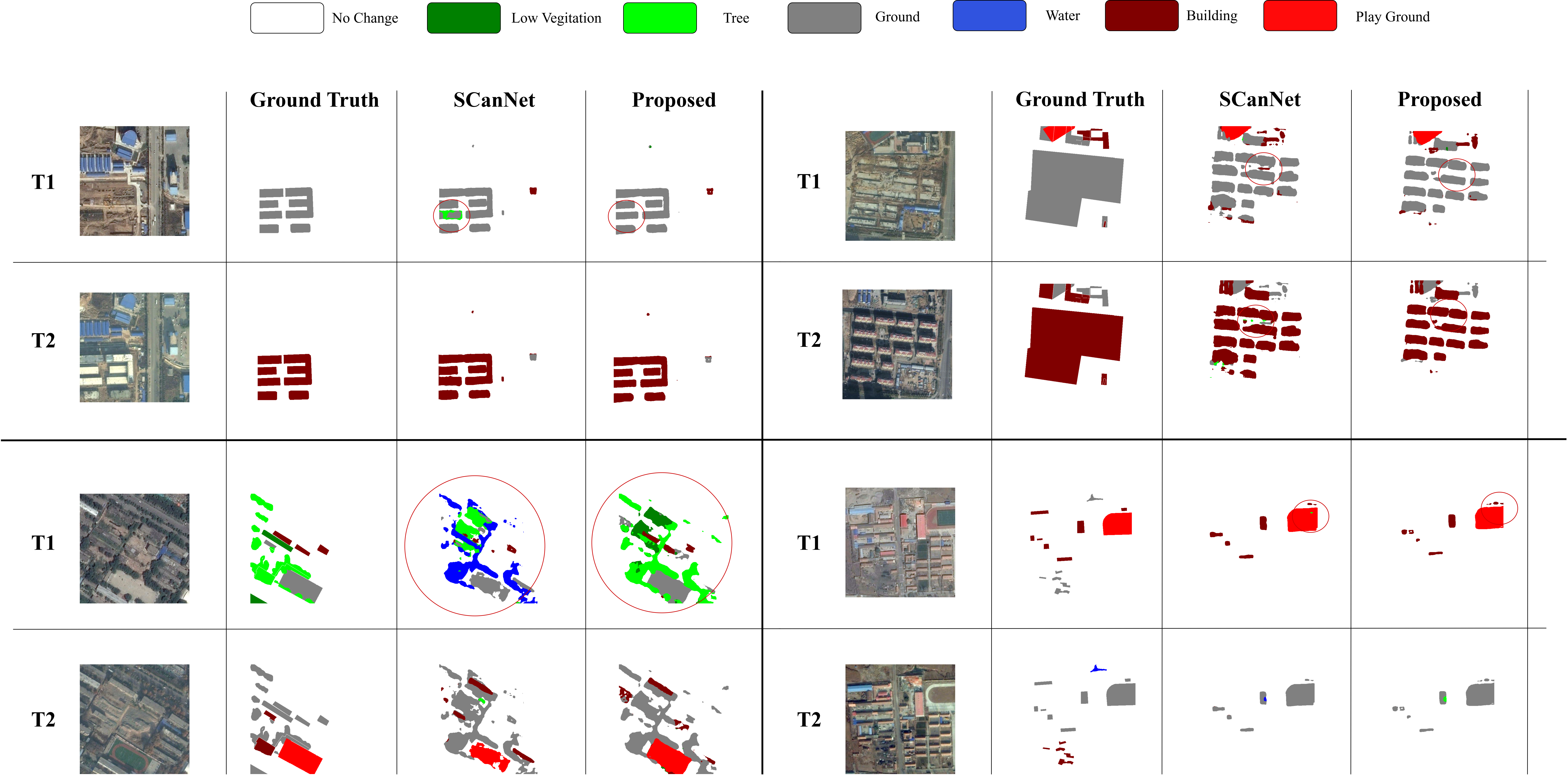}
    \caption{Semantic maps obtained by SCanNet Architecture and Proposed Architecture, Areas circled in red illustrates improvements achieved by the proposed architecture}
    \label{fig:enter-label}
\end{figure*}
\subsection{SECOND Dataset \cite{A_Shi_2022}}
The SECOND dataset comprises 4,662 pairs of 512$\times$512 aerial images with spatial resolutions ranging from 0.5 to 3 meters per pixel. It is annotated for SCD tasks in urban areas, including cities such as Chengdu, Shanghai, and Hangzhou in China. The dataset focuses on six primary land cover classes, non-vegetated ground surfaces, trees, low vegetation, water, buildings, and playgrounds, thereby capturing a wide spectrum of geographical changes. Each image pair is accompanied by corresponding land-cover maps and non-change masks, enabling precise differentiation between changed and unchanged regions, including those with intra-class variations. Following the experimental setup in \cite{J_Ding_2024}, we use 933 image pairs (1/5 of the dataset) as the test set and the remaining 3,729 pairs for training.

\subsection{Evaluation Metrics}
To assess the performance of our models, we employ four commonly used evaluation metrics in the literature, Overall Accuracy (OA), mean Intersection over Union (mIoU), Separated Kappa (SeK) coefficient, and the SCD-targeted F1 score (F\textsubscript{scd}). Formal definitions of these metrics are provided in \cite{J_Ding_2024, A_Shi_2022}.

\subsection{Training Settings}
All experiments were conducted using the PyTorch framework with a batch size of 6. Following the training setup in \cite{J_Ding_2024}, the initial learning rate($lr$) was set to 0.1 and adjusted over the first 50 epochs according to the polynomial decay schedule,
\begin{equation} lr = 0.1 \left(1 - \frac{\text{iterations}}{50} \right)^{1.5} \end{equation}

After 50 epochs, the best-performing model was selected. Optimization was performed using stochastic gradient descent (SGD) with Nesterov momentum.

\section{EXPERIMENTS AND DISCUSSION}

We conducted comprehensive ablation studies to quantitatively assess the impact of the proposed enhancements—CBAM and Dice loss—on a SECOND dataset. All models were trained with identical hyperparameters, and results were averaged over five independent runs to ensure statistical robustness. For qualitative evaluation, we compared our method against the original SCanNet using the pretrained model provided by the authors in \cite{J_Ding_2024}. Additionally, we performed a comparative analysis with SOTA models to further validate the effectiveness of our approach

\subsection{\textbf{Effect of CBAM}}

Table~\ref{tab:ablation} shows that integrating the CBAM into SCanNet leads to consistent improvements across all performance metrics. Specifically, the OA improves from 87.86\% to 87.98\%, F\textsubscript{scd} rises from 63.66\% to 63.81\%, mIoU from 73.42\% to 73.51\%, and SeK from 23.94\% to 24.11\%.

Although the numerical gains may appear modest, their consistency across multiple evaluation criteria underscores CBAM’s utility in enhancing representational quality. CBAM contributes by refining feature maps through two sequential attention mechanisms: channel attention, which emphasizes the most informative feature channels, and spatial attention, which focuses the model’s attention on the most relevant regions in the spatial domain\cite{C_Woo_2018}. These mechanisms enable the network to suppress irrelevant or misleading activations—such as artefacts from vegetation or shadows—while strengthening signals that correspond to semantically meaningful changes.

By integrating CBAM into the decoder branches, the model is better able to align low-level spatial detail with high-level semantic content. This leads to more precise segmentation boundaries and stronger detection of subtle changes. Notably, the improvements in F\textsubscript{scd} and SeK highlight CBAM’s capacity to enhance sensitivity to change regions while maintaining semantic consistency in unchanged areas.

In summary, CBAM improves semantic change detection by guiding the network to attend to both the “what” (important channels) and the “where” (important spatial locations), resulting in more robust and semantically coherent predictions. These findings demonstrate the value of attention-based refinement for overcoming challenges such as noisy inputs, small object classes, and class imbalance in remote sensing imagery.

\subsection{\textbf{Effect of Dice Loss}}
In remote sensing, semantic classes often exhibit imbalance, with changed or minority classes occupying small areas. Dice loss normalized by the sum of predictions and ground truths, focusing on the relative overlap, addresses this issue\cite{A_Zhou_2023,G_H._2017}.
Dice loss has minority class sensitivity and minority class resilience, which are ampliflications of the loss function when overlap is pure for rare classes with few positive pixels, and preventing large pixel count from dominating the loss for dominant classes respectively. These characteristics translate into measurable gains in our ablation: adding Dice on top of CBAM-SCanNet lifts Overall Accuracy from 87.98\% to 88.12 \%, $F_{scd}$ from 63.81 \% to 64.31 \%, mIoU from 73.51 \% to 73.63 \%, and SeK from 24.11 \% to 24.24 \% (Table \ref{tab:ablation}).

\subsection{\textbf{Qualitative Analysis}}
Across all four scenarios depicted in Figure \ref{fig:enter-label}, the proposed model consistently outperforms the original SCanNet by reducing false positives, recovering small or thin change regions, and producing sharper class boundaries.

In the top‐left panel, a single ground parcel has been levelled and a small outbuilding added. The baseline model correctly identifies the parcel but fragments the building’s thin roof edge and erroneously inserts a narrow vegetation strip at the boundary. By contrast, our model yields a continuous roof outline and no vegetation artifact, demonstrating both improved suppression of false positives and enhanced recall on sparse roof pixels.

In the top‐right panel, the post‐change image contains multiple narrow buildings. The baseline erroneously labels the inter‐building gaps as trees, whereas the proposed architecture correctly omits these false positives, faithfully preserving the building layout.

In the bottom‐left panel, the pre‐change image shows a tree stand that the baseline mistakenly predicts as water. Our model accurately classifies this region as vegetation. Moreover, in the corresponding post‐change image, it captures the sharp edges of the newly introduced playground area with high fidelity.

Finally, in the bottom‐right panel, the baseline produces spurious tree predictions within the playground, but our model successfully eliminates these artifacts. These qualitative improvements corroborate the quantitative gains reported in Table \ref{tab:ablation}.

\begin{table}[ht]
\centering
\caption{Comparison with the SOTA methods for SCD for SECOND dataset. \textcolor{red}{Red} indicates best, \textcolor{blue}{Blue} indicates second-best, and \textbf{Black} indicates third-best in each evaluation metrics}
\label{tab:sota_results}
\begin{tabular}{lcccc}
\toprule
\textbf{Method} & \textbf{OA(\%)} & \boldmath$F_{scd}$(\%) & \textbf{mIoU(\%)} & \textbf{SeK(\%)} \\
\midrule
ResNet‑GRU \cite{L_Mou_2019}       & 85.09 & 45.89 & 60.64 &  8.99 \\
ResNet‑LSTM \cite{L_Mou_2019}     & 86.91 & 57.05 & 67.27 & 16.14 \\
FC‑Siam‑conv. \cite{F_Caye_2018}    & 86.92 & 56.41 & 68.86 & 16.36 \\
FC‑Siam‑diff \cite{F_Caye_2018}     & 86.86 & 56.20 & 68.96 & 16.25 \\
HRSCD‑str.2 \cite{M_Caye_2019}      & 85.49 & 49.22 & 64.43 & 10.69 \\
HRSCD‑str.3 \cite{M_Caye_2019}       & 84.62 & 51.62 & 66.33 & 11.97 \\
HRSCD‑str.4 \cite{M_Caye_2019}      & 86.62 & 58.21 & 71.15 & 18.80 \\
SCDNet \cite{S_Peng_2021}           & 87.46 & 60.01 & 70.97 & 19.73 \\
SSCD‑l \cite{B_Ding_2022}           & 87.19 & 61.22 & 72.60 & 21.86 \\
Bi-SRNet \cite{B_Ding_2022}        & \textbf{87.84}  & \textbf{62.61}  & \textbf{73.41}  & \textbf{23.22}  \\
SMNet \cite{S_Niu_2023}            & 86.68  & 60.34  & 71.95  & 20.29  \\
TED \cite{J_Ding_2024}       & 87.39  & 61.56  & 72.79  & 22.17  \\
SCanNet\cite{J_Ding_2024}              & \textcolor{blue}{87.86}  & \textcolor{blue}{63.66}  & \textcolor{blue}{73.42}  & \textcolor{blue}{23.94}  \\
\midrule
\textbf{Proposed}
     & \textcolor{red}{88.12} & \textcolor{red}{64.31} & \textcolor{red}{73.63} & \textcolor{red}{24.25} \\
\bottomrule
\end{tabular}
\end{table}

\subsection{\textbf{Comparative Analysis}}
The proposed architecture outperforms all compared state-of-the-art semantic change detection methods on the SECOND benchmark, achieving the highest OA, $F_{scd}$, mIoU, and SeK among those listed in Table \ref{tab:sota_results}.

\section{Conclusion}
Our study demonstrates that integrating the CBAM and Dice loss into SCanNet significantly enhances the semantic change detection performance in remote sensing imagery. The CBAM effectively improves feature representation by sequentially applying channel and spatial attention, refining both channel importance and spatial accuracy. The Dice loss further addresses the prevalent issue of class imbalance, specifically enhancing the model's sensitivity to minority classes and fine-grained changes. Quantitative evaluations reveal consistent improvements across metrics such as Overall Accuracy, F1 score for semantic change detection (Fscd), mean Intersection over Union (mIoU), and Semantic edge-aware Kappa (SeK). Qualitative analyses reinforce these findings, illustrating clearer segmentation boundaries and improved detection of subtle changes. These enhancements collectively confirm the potential of attention mechanisms and specialized loss functions to address inherent challenges in semantic change detection tasks.

\bibliographystyle{unsrt}
\bibliography{references}

\begin{thebibliography}{10}

\bibitem{A_Zhu_2024}
Qiqi Zhu.
\newblock A review of multi-class change detection for satellite remote sensing imagery.
\newblock {\em Geo-spatial Information Science}, 27(1):1--15, Jan 2024.

\bibitem{C_Selvaraj_2022}
Sureshkumar~Nagarajan Rohini~Selvaraj.
\newblock Change detection techniques for a remote sensing application: An overview.
\newblock {\em Cognitive Systems and Signal Processing in Image Processing}, pages 129--143, 2022.

\bibitem{A_Zhou_2024}
Rufan Zhou.
\newblock A unified deep learning network for remote sensing image registration and change detection.
\newblock {\em IEEE Transactions on Geoscience and Remote Sensing}, 62:1--16, 2024.

\bibitem{C_Leichtle_2017}
Tobias Leichtle.
\newblock Class imbalance in unsupervised change detection – a diagnostic analysis from urban remote sensing.
\newblock {\em International Journal of Applied Earth Observation and Geoinformation}, 60:83--98, Aug 2017.

\bibitem{D_Bai_2023}
Ting Bai.
\newblock Deep learning for change detection in remote sensing: a review.
\newblock {\em Geo-spatial Information Science}, 26(3):262--288, Jul 2023.

\bibitem{U_Saha_2019}
Sudipan Saha.
\newblock Unsupervised deep change vector analysis for multiple-change detection in vhr images.
\newblock {\em IEEE Transactions on Geoscience and Remote Sensing}, 57(6):3677--3693, Jun 2019.

\bibitem{A_Bovolo_2008}
F.~Bovolo.
\newblock A novel approach to unsupervised change detection based on a semisupervised svm and a similarity measure.
\newblock {\em IEEE Transactions on Geoscience and Remote Sensing}, 46(7):2070--2082, Jul 2008.

\bibitem{A_IM_2005}
J~IM.
\newblock A change detection model based on neighborhood correlation image analysis and decision tree classification.
\newblock {\em Remote Sensing of Environment}, 99(3):326--340, Nov 2005.

\bibitem{G_Vemulapalli_2016}
Raviteja Vemulapalli.
\newblock Gaussian conditional random field network for semantic segmentation.
\newblock {\em 2016 IEEE Conference on Computer Vision and Pattern Recognition (CVPR)}, pages 3224--3233, Jun 2016.

\bibitem{C_Chen_2024}
Hongruixuan Chen.
\newblock Changemamba: Remote sensing change detection with spatiotemporal state space model.
\newblock {\em IEEE Transactions on Geoscience and Remote Sensing}, 62:1--20, 2024.

\bibitem{R_Li_2021}
Shujun Li.
\newblock Remote sensing image change detection based on fully convolutional network with pyramid attention.
\newblock {\em 2021 IEEE International Geoscience and Remote Sensing Symposium IGARSS}, pages 4352--4355, Jul 2021.

\bibitem{A_Jonasova_2023}
Eleonora~Jonasova Parelius.
\newblock A review of deep-learning methods for change detection in multispectral remote sensing images.
\newblock {\em Remote Sensing}, 15(8):2092, Apr 2023.

\bibitem{J_Ding_2024}
Lei Ding.
\newblock Joint spatio-temporal modeling for semantic change detection in remote sensing images.
\newblock {\em IEEE Transactions on Geoscience and Remote Sensing}, 62:1--14, 2024.

\bibitem{S_Liu_2021}
Ze~Liu.
\newblock Swin transformer: Hierarchical vision transformer using shifted windows.
\newblock {\em 2021 IEEE/CVF International Conference on Computer Vision (ICCV)}, pages 9992--10002, Oct 2021.

\bibitem{A_Yuan_2022}
Panli Yuan.
\newblock A transformer-based siamese network and an open optical dataset for semantic change detection of remote sensing images.
\newblock {\em International Journal of Digital Earth}, 15(1):1506--1525, Dec 2022.

\bibitem{C_Woo_2018}
Sanghyun Woo.
\newblock Cbam: Convolutional block attention module.
\newblock {\em Lecture Notes in Computer Science}, pages 3--19, 2018.

\bibitem{G_H._2017}
Carole~H. Sudre.
\newblock Generalised dice overlap as a deep learning loss function for highly unbalanced segmentations.
\newblock {\em Lecture Notes in Computer Science}, pages 240--248, 2017.

\bibitem{D_He_2016}
Kaiming He.
\newblock Deep residual learning for image recognition.
\newblock {\em 2016 IEEE Conference on Computer Vision and Pattern Recognition (CVPR)}, Jun 2016.

\bibitem{A_Shi_2022}
Qian Shi.
\newblock A deeply supervised attention metric-based network and an open aerial image dataset for remote sensing change detection.
\newblock {\em IEEE Transactions on Geoscience and Remote Sensing}, 60:1--16, 2022.

\bibitem{A_Zhou_2023}
Zheng Zhou.
\newblock A dynamic effective class balanced approach for remote sensing imagery semantic segmentation of imbalanced data.
\newblock {\em Remote Sensing}, 15(7):1768, Mar 2023.

\bibitem{L_Mou_2019}
Lichao Mou.
\newblock Learning spectral-spatial-temporal features via a recurrent convolutional neural network for change detection in multispectral imagery.
\newblock {\em IEEE Transactions on Geoscience and Remote Sensing}, 57(2):924--935, Feb 2019.

\bibitem{F_Caye_2018}
Rodrigo~Caye Daudt.
\newblock Fully convolutional siamese networks for change detection.
\newblock {\em 2018 25th IEEE International Conference on Image Processing (ICIP)}, pages 4063--4067, Oct 2018.

\bibitem{M_Caye_2019}
Rodrigo~Caye Daudt.
\newblock Multitask learning for large-scale semantic change detection.
\newblock {\em Computer Vision and Image Understanding}, 187:102783, Oct 2019.

\bibitem{S_Peng_2021}
Daifeng Peng.
\newblock Scdnet: A novel convolutional network for semantic change detection in high resolution optical remote sensing imagery.
\newblock {\em International Journal of Applied Earth Observation and Geoinformation}, 103:102465, Dec 2021.

\bibitem{B_Ding_2022}
Lei Ding.
\newblock Bi-temporal semantic reasoning for the semantic change detection in hr remote sensing images.
\newblock {\em IEEE Transactions on Geoscience and Remote Sensing}, 60:1--14, 2022.

\bibitem{S_Niu_2023}
Yiting Niu.
\newblock Smnet: Symmetric multi-task network for semantic change detection in remote sensing images based on cnn and transformer.
\newblock {\em Remote Sensing}, 15(4):949, Feb 2023.

\end{thebibliography}

\end{document}